\documentclass[format=acmsmall, review=false, screen=true]{acmart}

\usepackage{booktabs} 

\usepackage{balance}    
\usepackage{enumerate}
\usepackage[inline]{enumitem}

\usepackage{filecontents}

\usepackage{xcolor,colortbl}
\definecolor{Gray}{gray}{0.85}
\usepackage{color,soul}
\usepackage{multirow}

\begin{document}

\title[Intelligent Notification Systems: A Survey of the State of the Art and Research Challenges] {Intelligent Notification Systems: A Survey of the State of the Art and Research Challenges}
  
\author{Abhinav Mehrotra}
\affiliation{University College London, UK}
\author{Mirco Musolesi}
\affiliation{University College London and Alan Turing Institute, UK}

\begin{abstract}
Notifications provide a unique mechanism for increasing the effectiveness of real-time information delivery systems. However, notifications that demand users' attention at inopportune moments are more likely to have adverse effects and might become a cause of potential disruption rather than proving beneficial to users. In order to address these challenges a variety of intelligent notification mechanisms based on monitoring and learning users' behavior have been proposed. The goal of such mechanisms is maximizing users' receptivity to the delivered information by automatically inferring the right time and the right context for sending a certain type of information.

This article presents an overview of the current state of the art in the area of intelligent notification mechanisms that relies on the awareness of users' context and preferences. More specifically, we first present a survey of studies focusing on understanding and modeling users' interruptibility and receptivity to notifications from desktops and mobile devices. Then, we discuss the existing challenges and opportunities in developing mechanisms for intelligent notification systems in a variety of application scenarios.

\end{abstract}

%
%


%
%



\maketitle

\renewcommand{\shortauthors}{A. Mehrotra and M. Musolesi}

\section{Introduction}
\label{sec:introduction}

Mobile phones represent an essential element of our lives by assisting us with multifarious day-to-day activities. At the same time, they are always connected to the Internet. For these reasons, they represent a unique platform for receiving or fetching information anytime and anywhere. For example, they can be used to keep users aware of information channels through mobile apps such as Skype, Hangouts, Facebook, Twitter and Gmail. 

In order to ensure real-time awareness of users about the delivered information, mobile operating systems rely on \textit{notifications} that steer users' attention towards the delivered information through audio, visual and haptic signals. This is indeed in contrast with the traditional paradigm of \textit{pull-based} information retrieval and delivery in which the user has to initiate a request for the transmission of information. Notifications are the cornerstone of \textit{push-based} information delivery via mobile phones as they allow applications to harness the opportunity of steering users' attention towards the delivered information in order to maximize its effectiveness. Mobile notifications are presented in a unified fashion by almost all mobile operating systems. Usually, in the current implementations, notifications from all applications are listed in a notification bar at the top of a phone's screen. In order to provide a quick glimpse of the delivered information to the users, they present a brief summary including the identity of the sender, a brief description or summary of the content of the notifications or the event that trigger them, and time of delivery.

Mobile notifications are generated by humans as well as machines. The former are generated by recipient's social connections generally through chat and email applications for instantiating communication between two or more persons, whereas, in the latter case, messages are generated in an automatic fashion by the system processes or the native applications, such as system monitoring utilities, scheduled reminders and promotional advertisements. Nowadays, since a variety of sophisticated sensors are embedded in phones, the data generated therefrom is exploited by sophisticated applications for not just improving usability but also for proactively signaling users about the occurrence of events that are associated with their context. Some common example of context-based notifications are collocation-based advertisements~\cite{Aalto2004, Foth2010} and context-based suggestions~\cite{Lane2011b, WalkSafe2012}.

\begin{figure}[t]
\centering
\includegraphics[width=0.45\textwidth]{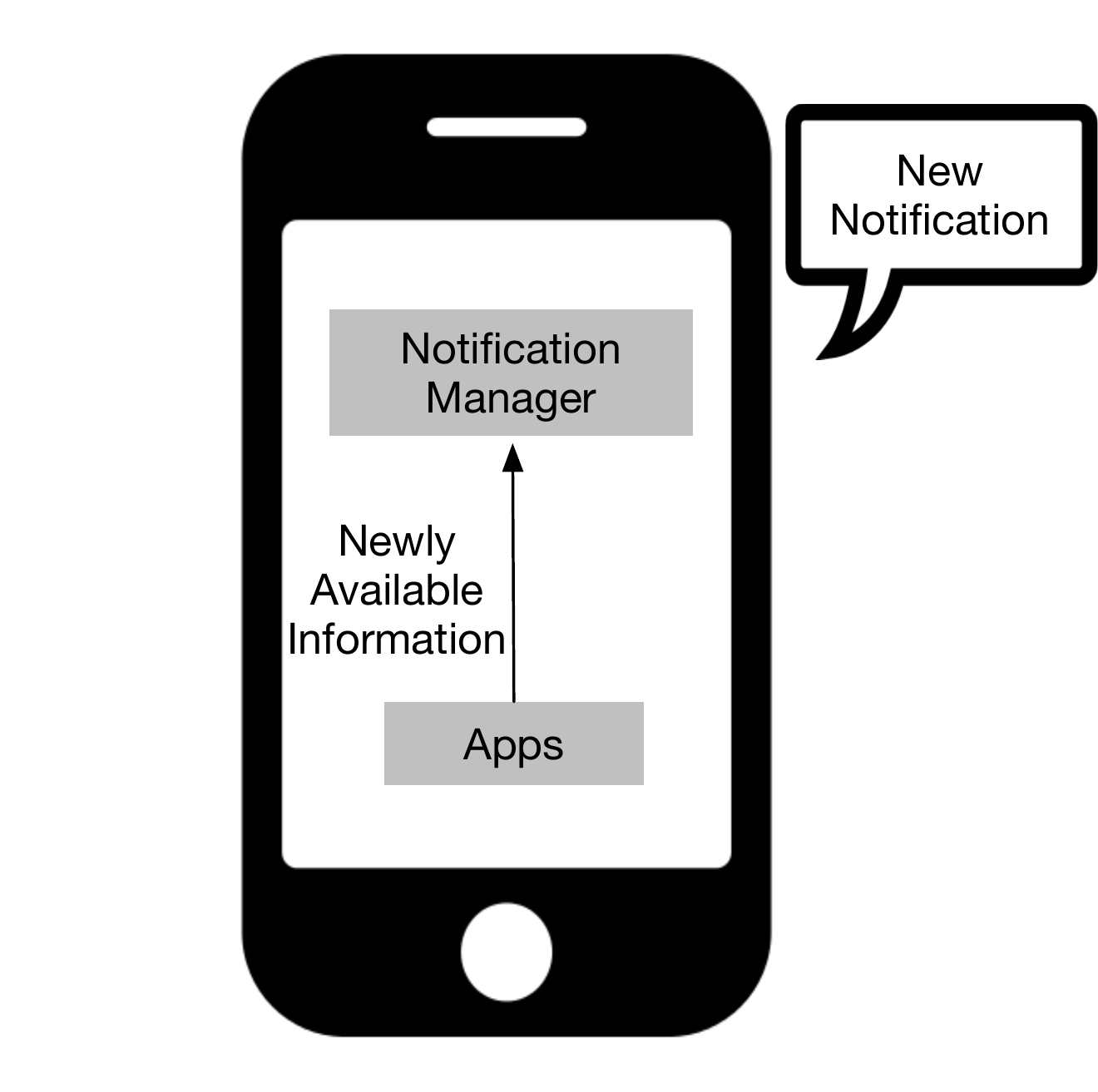}
\caption{Information delivery through push notifications on mobile phones.} 
\label{fig:notification}
\end{figure}

Push notifications (as shown in Figure~\ref{fig:notification}) were introduced in mobile phones to keep users free from constantly checking for (i.e., pulling) new information, as they signal users on the availability of any newly arrived information. However, people receive numerous notifications arriving autonomously at anytime during the day through their mobile apps~\cite{Sahami2014LargeScale, Mehrotra2015NotifyMe}. Such services provide a clear benefit to users as they facilitate task switching and keep users aware of a number of information channels in an effortless manner. However, at the same time, these notifications are often triggered at inappropriate times as they do not have any knowledge about recipients' situation. Psychological studies have found that notifications arriving at unsuitable moments often become a cause of disruption for the on-going task~\cite{Miyata1986Psychological, Zijlstra1999}. Notifications delivered at the wrong time can adversely affect the current primary task's execution~\cite{ailey2000Interruptions, Cutrell2001Notification, Czerwinski2000Instant, Monk2002Attentional} and users' affective state~\cite{Adamczyk2004If, Bailey2006Need}.


On the other hand, a previous study has found that in order to not miss any newly available information that is deemed important, people are willing to tolerate mobile interruptions~\cite{Iqbal2010Notifications}. However, various mobile applications exploit users' willingness by triggering a large number of potentially unwanted notifications~\cite{Mehrotra2015NotifyMe}. At the same time, some studies have demonstrated that not all notifications are accepted by users as their receptivity is dependent on on the type and sender of information being delivered~\cite{Mehrotra2015NotifyMe, Mehrotra2016MyPhoneAndMe}. For this reason, notifications that are uninteresting and irrelevant are mostly dismissed (i.e., swipe away without clicking) by users~\cite{Fischer2010Effects, Sahami2014LargeScale}. Some examples include new game invites, app updates, predictive suggestions by recommendation systems and marketing messages. Furthermore, continual trigger of notifications at inappropriate time and context becomes a potential cause of annoyance for users. This might lead them to uninstall the corresponding applications~\cite{Felt2012Problems, Sahami2014LargeScale}. In order to provide an in-depth discussion of these issues, we devote a part of the survey to the issues concerning the cost of interruptions for both users and the service providers.

In the recent years, studies have focused on exploiting the contextual information (captured through mobile sensors) for understanding users' notification-interaction behavior~\cite{Sahami2014LargeScale, Mehrotra2016MyPhoneAndMe}. Indeed, the interaction of users with notifications is conditional on various contextual dimensions, which might be partially captured by means of the sensors embedded in the phones. However, by monitoring users' interaction with notifications, researchers have been developing a variety of  interruptibility management mechanisms for both desktop and mobile environments. More specifically, these studies have been focused on: 
\begin{enumerate*}
\item understanding factors associated with users' interruptibility and receptivity to notifications;
\item inferring opportune moments and contextual conditions for notification delivery;
\item identifying and filtering notifications that are deemed uninteresting or irrelevant for users.
\end{enumerate*}
In this survey we discuss previous efforts and state-of-the-art in interruptibility management system design for both desktop and mobile environments separately. We present a critical discussion of these approaches and highlight the current challenges and opportunities.

More specifically, this paper is structured as follows. In Section~\ref{sec:interruptions}, we frame the problem by examining possible definitions of \textit{interruptions} based on past and current theories, and give an overview of their types and sources. In Section~\ref{sec:cost} and~\ref{sec:differences}, we discuss the key findings of studies concerning the cost of interruptions and individual differences in perceiving them in detail. 
In Section~\ref{sec:im} we discuss the definition of \textit{interruptibility management}. We then present a comprehensive coverage of practical approaches for dealing with interruptions in desktop (Section~\ref{sec:im_desktop}) and mobile (Section~\ref{sec:im_mobile}) settings. 
Finally, in Section~\ref{sec:discussion} we highlight the limitations of previous approaches and discuss a series of potential research opportunities in this area.

\section{Interruptions}
\label{sec:interruptions}

\subsection{Definitions of Interruptions from Different Research Fields}

The concept of \textit{interruption} has been defined and interpreted in different ways by researchers working in different communities. For example, in linguistics, an interruption has been defined as:\\
``\textit{A piece of discourse that breaks the flow of the preceding discourse. An interruption is in some way distinct from the rest of the preceding discourse; after the break for the interruption, the discourse returns to the interrupted piece of discourse~\cite{Grosz1986Attention}.}''
\\ \\
An interruption is defined as follows in psychology:\\
``\textit{An event that breaks the coherence of an ongoing task and blocks its further flow. However, people can resume the primary task that has been interrupted once the interruption is removed~\cite{McFarlane2002HCIDesign}.}''
\\ \\ 
Instead, in computer science, it is defined as:\\
``\textit{An event prompting transition and reallocation of attention focus from a task to the notification~\cite{McCrickard2003NotificationSystem}.}'' 
\\

\subsection{Types of Interruptions}

Interruptions are pervasive in nature. As described by Miyata and Norman~\cite{Miyata1986Psychological}, in our everyday life we receive numerous interruptions that are both internal and external. 
An overview of the definitions of these two types of interruptions are as follows: 

\subsubsection{Internal Interruptions} 
An internal interruption occurs due to our own background thought process. More specifically, such interruptions are actions performed by people themselves that lead to break their focus of conscious attention to perform another activity. 

\subsubsection{External Interruptions} 
An external interruption is caused by the arrival of an event around a user. Communication through computing devices or in-person is the fundamental source of external interruptions. 
External interruptions can be further divided into two classes depending on the relevance of sources with the primary task:
\begin{itemize}
\item{\textbf{Implicit Interruptions}: 
These are the interruptions that arrive from some process of a primary task. Such interruptions are mostly relevant to the current task, for example, an error message from the application which a person is interacting with. }

\item{\textbf{Explicit Interruptions}: 
These are the interruptions that arrive from a process that does not belong to the ongoing task. Such an interruption causes an expected task switch from the current activity to a newly introduced activity, for example, the arrival of a chat message while a person is interacting with a text editor application. }
\end{itemize}

\subsection{Definition of Interruptions in Context of this Survey}

Previous theories about interruptions in different domains enable us to gain a deeper insights about the meaning of interruptions. Starting from this previous work from different communities, we first derive a generic theoretical construct for providing a definition of \textit{interruptions} and framing the problem of \textit{interruptibility}. The objective of this article is to focus on external interruptions through desktop and mobile devices without considering internal interruptions. Therefore, we define an interruption as \textit{an unanticipated event that comes through a communication medium and has a potential to instigate a task switch and break the flow of the primary activity by capturing users' attention through visual, auditory, or haptic cues}.

\subsection{Sources of Interruptions}
Given the focus on external interruptions, in this section we discuss the sources of such interruptions and the ways in which users handle them. 

\subsubsection{Interruptions in Human-Human Discourse} 
In a human-human communication environment, when an interaction is initiated by a person (i.e., speaker), the listener generally gives feedback about the failure or success of the initiated communication~\cite{Clark1989}. Such feedback is merely a brief reaction through eye contacts, head nods or  voice response. This acknowledgement informs the speaker whether the listener is welcoming and attending the communication or not. 

As suggested by Clark and Schaefer~\cite{Clark1989}, ``\textit{[f]or people to contribute to discourse, they must utter the right sentence at the right time}''. The physical presence of a person enables the speaker to understand the right moment to initiate communication. However, the speaker does not usually have a reasonable understanding of the listener's cognitive situation. Consequently, people sometimes initiate communication at wrong moments, which often results in causing interruptions. 

As suggested by Clark~\cite{Clark1996UsingLanguage}, a listener responds to such interruptions in four possible ways:
\begin{enumerate}
\item responding to interruptions immediately.
\item acknowledging and agreeing to handle it later.
\item explicitly refusing to handle it. 
\item implicitly refusing to handle it by not providing an acknowledgement to the interruption.
\end{enumerate}

On the other hand, once the conversation begins, it does not always go error-free. As argued by Sacks et al. in~\cite{Sacks1974}, the turn-taking during the conversation is itself vulnerable to error. When a person speaks the other person listens, but sometimes unintentionally both start talking simultaneously and people try to coordinate their conversation in an appropriate way. However, this can be categorized as an implicit interruption (discussed earlier in this section), which is not in the scope of this article.

\subsubsection{Interruptions in Human-Computer Interaction}

Personal computing devices such as desktops, laptops and mobile phones, offer a great value to users by facilitating multifarious, informative and computational functionalities salient to their daily requirements. However, the provision of a platform that runs multiple applications simultaneously, which are delivering various types of information from different channels, often leads to an environment that distracts users from their primary task\footnote{Here a primary task can be any operation which the user is currently performing, which might or might not include interaction with a computing device.}.
This is due to the fact that the information delivered to the users is often not relevant to their primary task, which leads them to switch their attention from the application in focus towards the application being executed in the background that delivered the information. Moreover, applications leverage notifications in order to trigger alerts that try to gain users' attention towards the delivered information. Therefore, even though people try to ignore all interruptions and continue focusing on their primary task, they still receive cues about the newly delivered information, which might cause information overload~\cite{Speier1999}.  

In 1997, by looking at the trend towards the development of ``intelligent'' computer system and technologies, McFarlane envisioned the hazards from such intelligent systems competing for users' attention~\cite{Mcfarlane1997}. McFarlane, for the first time, investigated the strategies for counteracting interruptions caused by intelligent computer systems. He built a taxonomy based on theoretical constructs that are relevant to interruptions. This taxonomy identifies the following eight descriptive aspects of human interruption: 
\begin{enumerate}[label=(\roman*)]
\item \textit{Source of interruption}: who triggered the interruption.
\item \textit{Characteristics of the user being interrupted}: receiver's perspective for getting interrupted.
\item \textit{Coordination method}: approach used for determining the moment to trigger interruption based on users' response. 
\item \textit{Meaning of interruption}: what the interruption is about.
\item \textit{Method of expression}: design aspect of the interruption.
\item \textit{Channel of conveyance}: medium of receiving the interruption.
\item \textit{Human activity changed by interruptions}: internal or external change in the recipient's conscious and physical activity. 
\item \textit{Effect of interruption}: impact of interruption on an ongoing task and the user.
\end{enumerate}

Instead of limiting the scope of his work to HCI, McFarlane built this taxonomy from an interdisciplinary perspective drawing upon the theories for human interruption discussed in the literature from many different domains. Each dimension of the taxonomy describes a unique aspect of human interruptibility. 

In 1999, Latorella proposed an Interruption Management Stage Model (IMSM) that describes information processing stages on receiving interruptions~\cite{Latorella1999IMSM}. This model can be used to study information processing by humans and to identify the effects of interruptions in different stages of information processing. The model was designed with an assumption that recipients are engaged with an ongoing task with which they are familiar and that it can be resumed at any point. It comprises of three stages: 
\begin{enumerate}[label=(\roman*)]
\item \textit{Interruption detection}: when a user is engaged with the primary task, a salient alert is required in order to initiate an interruption.
\item \textit{Interruption interpretation}: on detection of an interruption, the user's attention is directed towards the interruption for further processing in order to interpret the requirements of the interrupting task. 
\item \textit{Interruption integration}: in this final stage, the user integrates the interruption with the primary task by immediate or scheduled tasks switching.   
\end{enumerate}

In 2002, McFarlane and Latorella investigated users' behavior on receiving interruption alerts (i.e., the first stage of IMSM model)~\cite{McFarlane2002HCIDesign}. They argued that user's response to computer generated interruptions is similar to the response to interruptions during human dialogue as proposed by Clark~\cite{Clark1996UsingLanguage}. However, they suggested that Clark only considered the user's response for detected interruptions; indeed, in the human-computer interaction setting these can also go undetected. Therefore, undetected interruptions might represent additional aspects of user response, which should be considered when building interruption management system for computing environments. 

McFarlane and Latorella proposed the following five key responses of users to an interruption arriving during the process of human-computer interaction:
\begin{enumerate}[label=(\roman*)] 
\item \textit{Oblivious dismissal}: the interruption goes unnoticed by the user and, thus, it is not performed;  
\item \textit{Unintentional dismissal}: the interruption is not performed as it is not interpreted to the user;
\item \textit{Intentional dismissal}: the user explicitly decides not to handle the interruption;
\item \textit{Preemptive integration}: the interruption is handled immediately and the ongoing task is resumed after finishing the interrupting task;
\item \textit{Intentional integration}: the interruption is relevant to the ongoing task and the user integrates it with the ongoing task. 
\end{enumerate}

\section{Cost of Interruption}
\label{sec:cost}

\begin{table}[t]
\small
\centering 
\begin{tabular}{ | p{0.3\textwidth} p{0.65\textwidth} | } 
\hline 
\rowcolor{Gray}
\textbf{Study Type} & \textbf{Key Findings} \\ [0.05ex] 
\hline 

\multirow{4}{4cm}{Impact on Memory} & People have selective memory associated to the interruptions they receive~\cite{Zeigarnik1927}.\\
	& People possess a stronger memory representation of an uninterrupted task as compared to an interrupted task~\cite{Edwards1998Interruption, Gillie1989}.\\ 
\hline

\multirow{9}{4cm}{Relationship with Ongoing Task} & Interruptions could have an adverse effect on the completion time and errors made while performing a complex computing task compared to a simple task~\cite{kreifeldt1981interruption, Gillie1989}.\\
	& People perceive less disruption if the interruption is highly relevant to the current task~\cite{Czerwinski2000a}.\\ 
	& Amount of disruption perceived also linked to the mental load of a user on the arrival of an interruption~\cite{Bailey2000Interruptions, Bailey2001Interruptions}. \\
	& People perceive varying level of disruption while performing different sub-tasks~\cite{Czerwinski2000Instant, Cutrell2001Notification}. \\
\hline

\multirow{6}{4cm}{Relationship with Users' Emotional State} & People's emotion and well-being are negatively impacted by interruptions~\cite{Zijlstra1999}.\\
	& People experience annoyance and anxiousness on arrival of an interruption~\cite{Bailey2001Interruptions, Adamczyk2004If}. \\
	& Interruptions coming from mobile phones cause lack of attention and hyperactivity symptoms in users~\cite{Kushlev2016Silence}. \\
\hline

\multirow{2}{4cm}{Impact on User Experience} & Complex interfaces make it difficult for users to handle interruptions~\cite{kreifeldt1981interruption}.\\\hline

\end{tabular}
\caption{Studies in the area of understanding the cost of interruptions.} 
\label{table:cost} 
\end{table}

Interruptions are an inevitable part of our everyday life as it is hard to get through the entire day without being interrupted. As suggested by Zabelina et al. in~\cite{Zabelina2015Creativity}, people are sensitive to their surroundings and they receive more information through interruptions, which might help them in their everyday tasks and even boost their creativity. This represents a positive aspect of interruptions. At the same time, numerous studies~\cite{Edwards1998Interruption, Czerwinski2000a, Bailey2000Interruptions, Cutrell2001Notification} have demonstrated that interruptions have a detrimental effect on users' memory, emotional and affective states, and ongoing task execution. In this section we discuss the cost associated with the arrival of interruptions at inopportune moments. The summary of this section is presented in Table~\ref{table:cost}.

\subsection{Impact on Memory}
In 1927, Zeigarnik performed a classic psychological study~\cite{Zeigarnik1927} (as cited in~\cite{baddeley1976psychology}) with the goal of examining the mechanisms of retrospective remembering with and without interruptions. In this study the participants were given a series of practical tasks, for instance sketching a vase and putting beads on a string. Some tasks were interrupted and others were carried out  without any interruption. Tasks could be performed in any order by participants. It was possible to switch to another task without completing the ongoing task, which could be taken up later. On the completion of all tasks, they were asked to do a recall test. The results of this study demonstrate that people can recall the content of interrupted tasks better compared to the case of uninterrupted tasks, which might indicate that people have selective memory associated to the interruptions they receive. This observed behavior is referred to as the Zeigarnik Effect. 

Although the Zeigarnik Effect suggests that interruptions are useful for retrospective memory, many other applied studies have argued that interruptions have an adverse impact on memory~\cite{Dix1993HCI, Gillie1989, Edwards1998Interruption, Cutrell2001Notification}. In particular, Dix et al. argued that humans can memorize only a limited list of tasks they have been engaged in due to the nature of their cognitive abilities~\cite{Dix1993HCI}. Moreover, if interrupted during a task, humans are likely to lose track of what they were doing. Following these suggestions, Edwards and Gronlund~\cite{Edwards1998Interruption} conducted an experiment to investigate the memory representation for the primary task after handling an interruption. Their study was orthogonal to the Zeigarnik Effect experiment as in the latter participants were not asked to resume or to recall where they left the primary task on arrival of the interruption. Through their experiment, Edwards and Gronlund demonstrated that people possess a stronger memory representation of an uninterrupted task as compared to an interrupted task. Moreover, they showed that people tend to need a certain amount of time before resuming back to the primary task after an interruption.

\subsection{Impact on Ongoing Task Performance} 

In 1981, Kreifeldt and McCarthy~\cite{kreifeldt1981interruption} argued that interruptions could have an adverse effect on the completion time and errors made while performing a computing task. They demonstrated that people perceive more disruption and become more prone to make errors on getting interrupted while performing a complex task compared to a simple task. 
Later in 1989, through a series of experiments, Gillie and Broadbent~\cite{Gillie1989} investigated the impact on the primary task by three aspects of an interruption: (i) length; (ii) similarity with the primary task; (iii) the action required to handle it. Their results show that people feel distracted when interruptions share characteristics with the ongoing task or if they are limited to complex tasks but the length of an interruption does not make it disruptive. However, these findings are not aligned with those of Czerwinski et al.~\cite{Czerwinski2000a} who demonstrated that people perceive less disruption if the interruption is highly relevant to the current task. 

In~\cite{Bailey2000Interruptions} Bailey at al. studied whether the performance of an ongoing task is influenced by interruptions. Their experiment utilized six types of web-based task: addition of numbers, counting of items, comprehension of images, comprehension  of written text, registration and selection. Participants were interrupted when they were approximately in the midway to completion of each task. They were presented with a news report or an investment decision as an interruption. Their findings show that: (i) people perform interrupted tasks slower compared to non-interrupted tasks; and (ii) the amount of disruption perceived depends on the type of ongoing task. Later, in another study~\cite{Bailey2001Interruptions}, Bailey's et al. demonstrated that the amount of disruption perceived also depends on the mental load of a user on the arrival of an interruption. 

Czerwinski et al. in~\cite{Czerwinski2000Instant} studied the impact of interruptions while performing different types of sub-tasks. Their results show that people perceive varying levels of disruption while performing different sub-tasks. They proposed that deferring interruptions until a new subtask is detected could also reduce the perceived disruption. These findings were extended by Cutrell et al.~\cite{Cutrell2001Notification} to investigate the effects of instant messaging on different types of computing tasks. They found that the disruptiveness perceived is higher when users are engaged with tasks that require their attention.

\subsection{Impact on Users' Emotional State}

In~\cite{Zijlstra1999} Zijlstra et al. for the first time studied the effect of interruptions on users' psychological state. More specifically, they investigated whether interruptions produce an adverse effect on users' emotions and well-being, and raise their activeness level. They conducted a series of experiments by creating a simulated office environment for performing realistic text editing tasks. Their findings suggest that users' emotion and well-being are negatively impacted by interruptions, but they do not affect the activeness level.

In 2001, Bailey at al. investigated the effects of interruption on users' annoyance and anxiety levels~\cite{Bailey2001Interruptions}. Through a series of experiments, they demonstrated that people experience annoyance on arrival of an interruption. The annoyance level experienced by users depends on the type of ongoing task, but not on the type of the interruption task. They also show that the increase in users' anxiety level is higher when they receive interruptions during a primary task as compared to the arrival of interruption on completion of the primary task. 

In another study~\cite{Adamczyk2004If}, Adamczyk and Bailey investigated the impact on users' emotional state by interruptions arriving at particular times during task execution. In their experiment, participants were asked to perform tasks (such as text editing, searching and watching video) and a periodic news alert was triggered as an interruption. Their findings show that users experience annoyance and frustration on receiving interruptions. Moreover, interruptions arriving at different moments have a varying impact on users' emotional state.

In a study concerning mobile notifications~\cite{Kushlev2016Silence}, Kushlev et al. investigated whether interruptions coming from mobile phones cause lack of attention and symptoms linked to Attention Deficit Hyperactivity Disorder (ADHD). They asked participants to maximize interruptions by turning on their phones' notification alerts and trying to mostly be within the reach of their phone. Later, participants were asked to minimize interruptions by turning off their phones' notification alerts and trying to stay away from their phones. Their results show that people reported higher levels of hyperactivity and distraction during the first phase of the experiments. This suggests that by simply adjusting existing phone settings people can reduce inattention and hyperactivity levels.

\subsection{Impact on User Experience}

\indent 
\indent
\textit{``The most profound technologies are those that disappear. They weave themselves into the fabric of everyday life until they are indistinguishable from it.'' -- Mark Weiser~\cite{Weiser1991}. }

\noindent
The above quote from Mark Weiser's seminal paper~\cite{Weiser1991} provide a clear picture about his vision for ubiquitous computing. His goal was to design an environment with embedded unobtrusive computing and communication capabilities that can blend with users' day-to-day life. 
The two key aspects of his vision were: 
(i) effective use of the environment in order to fuse technology with it;
(ii) making the technology disappear in the environment. 

The second aspect of his vision focuses on the user experience and suggests making technology disappear from the user's consciousness. Another classic paper of Weiser~\cite{Weiser1997CalmTechnology} describes the disappearing technologies as \textit{calm computing}. In this paper he suggested that the technology should enable the seamless provision of information to users without demanding their focus and attention. However, interruptions have the potential to take mobile technology and Mark Weiser's vision far apart because they not only create potential information overload but also demand user attention. 


As suggested by Mark Weiser~\cite{Weiser1991}, technology should be ``transparent'' to users so that they should not notice that they are interacting with computing devices. In one of the first works in this area, Kreifeldt and McCarthy compared the design of user interfaces in order to reduce the effects of interruptions~\cite{kreifeldt1981interruption}. 
Their results suggest that the interaction design plays a role in affecting users' ability to successfully resume the interrupted task. More specifically, their findings suggest that complex interfaces make it difficult to handle interruptions. 



\section{Individual Differences in Perceiving Interruptions}
\label{sec:differences}

As defined earlier, interruptions are unanticipated events that fragment the flow of execution of an ongoing task by demanding users to switch their attention to interrupting tasks. 
This leads to a multitasking environment for users. Humans naturally have skills to handle interruptions and adapt to a multitasking environment; however, they do show individual differences in their ability to perform tasks in such setting~\cite{Atkinson1953, Weiner1965, Braune1986, Joslyn1998}.

In 1953, Atkinson investigated the role of people's motivation to recall the completed and interrupted tasks~\cite{Atkinson1953}. The study demonstrates that highly motivated users are likely to recall interrupted tasks better as compared to recalling uninterrupted tasks. On the other hand, less motivated people show a tendency to recall completed tasks better than recalling the interrupted task. Overall, his findings suggest that the ability to recall the primary task after handling interruptions varies across people. A few years later, similar findings were reported by Bernard Weiner~\cite{Weiner1965}.

Joslyn and Hunt proposed ``The Puzzle Game'' -- an empirically validated test that can quantify the performance of users for multitasking~\cite{Joslyn1998}. 
Through a series of experiments, the authors showed that the ability to make rapid decisions for task switching is not the same for everyone. They suggested that people who are good in making quick decisions can be identified through testing their psychological characteristics, which can be captured by their puzzle game. 

Brause and Wickens in~\cite{Braune1986} investigated the individual differences in sharing time between tasks in a multitasking environment. Their analysis show that there are differences in time-sharing ability of individuals; these are linked to their potential to process information. Moreover, their findings suggest that people have different strategies for time-sharing, which introduces differences in individuals' multitasking ability.

Moreover, some studies have demonstrated that people show significant differences in their cognitive style with respect to multitasking~\cite{Husain1987, Cabon1990Interruption}. Studies have demonstrated that the performance in carrying out an interrupted task is affected by users' anxiety~\cite{Husain1987} and arousal~\cite{Cabon1990Interruption} levels. However, the level of anxiety and arousal varies across people~\cite{Husain1987, Cabon1990Interruption}.

\section{Interruptibility Management}
\label{sec:im}

\begin{figure}[t]
\centering
\includegraphics[width=0.65\textwidth]{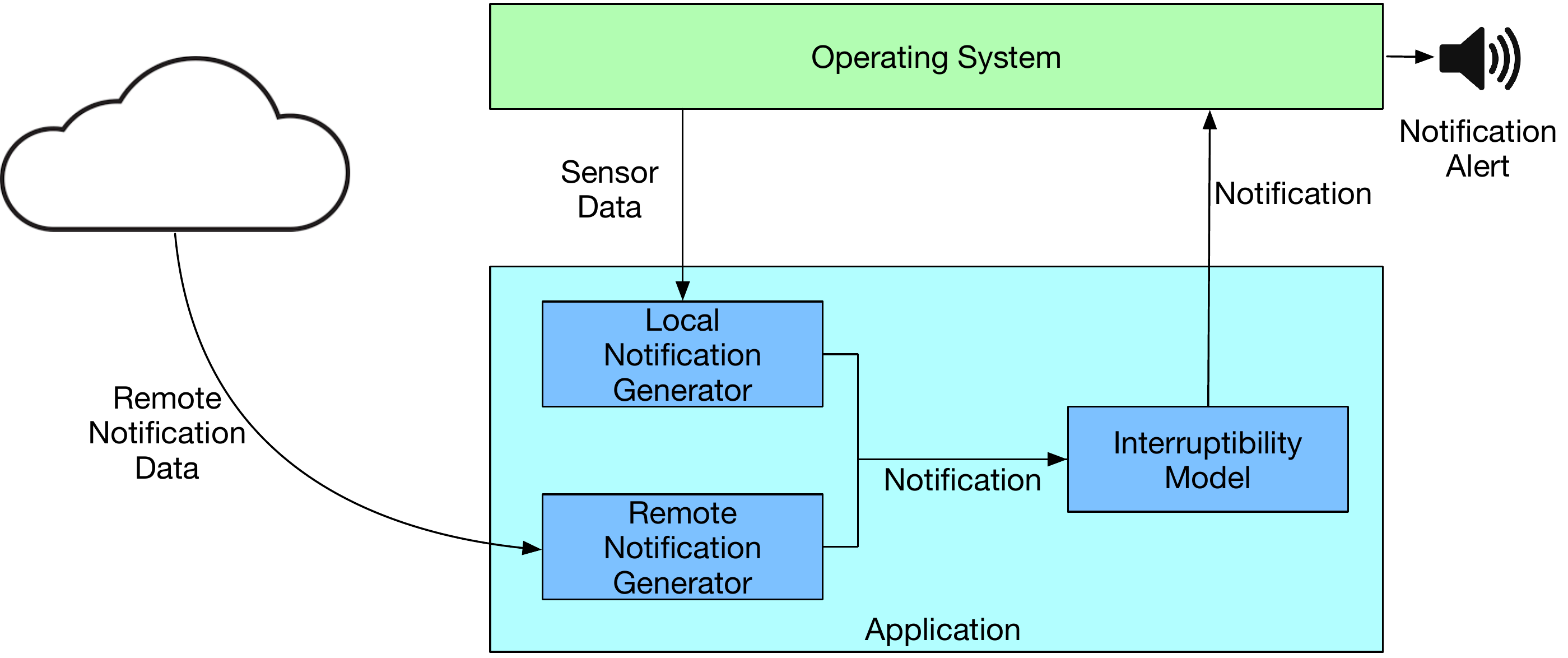}
\caption{Architecture of an interruptibility management mechanism.} 
\label{fig:architecture}
\end{figure}

\subsection{Definition}
Information delivered through computing devices often arrive at inopportune moments. This might adversely affect our ongoing tasks and psychological states. \textit{Interruption management} is a process that combines technology, practices and policies to build solutions for controlling interruptions from seeking users' attention at inopportune moments. The key objective of an interruptibility management system is to help users to effectively perform their primary task and make computing devices calm by unobtrusively mediating interruptions~\cite{Iqbal2005Towards}.

\begin{figure}[t]
\centering
\includegraphics[width=0.85\textwidth]{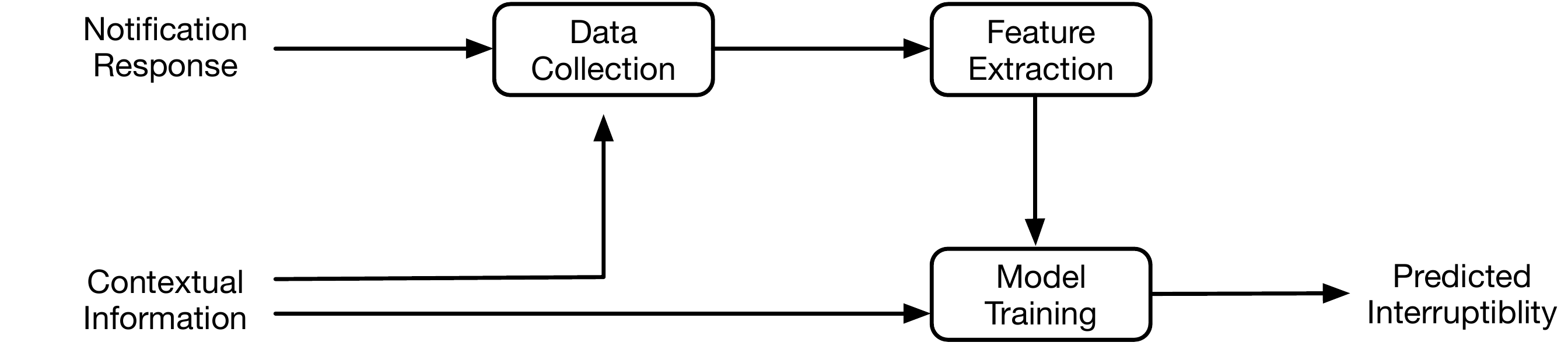}
\caption{Architecture of an interruptibility management mechanism.} 
\label{fig:model}
\end{figure}

In general, Figure~\ref{fig:architecture} presents the architecture of an interruptibility management mechanism entailed in an app. It relies on an \textit{interruptibility model} in order to handle interruptions from both local and remote notifications. Various approaches for constructing interruptibility models are discussed later in this section. A high level overview of the process to build these models is shown in Figure~\ref{fig:model}, which consists of three steps: data collection, feature construction and model training.

\subsection{Attentiveness and Receptivity to Interruption}

In~\cite{Pielot2014Didnt} \textit{attentiveness} is defined as the amount of attention paid by users towards their computing device for a newly available interruption task. However, attentiveness does not consider the response of the user to the interruption, which can be either positive (i.e., the interruption is accepted) or negative (i.e., the interruption is dismissed). On attending an interruption users get subtle clues about different features of interruptions such as a brief description of the content, sender and urgency of the interrupting task, which helps them to decide whether to click or decline those interruptions.

On the other hand, \textit{receptivity} is defined as the process of making a decision about the way in which the user is willing to respond to an interruption by analyzing its clues. In~\cite{Fischer2010Effects} Fischer at el. argue that users' receptivity to an interruption not only encompasses their reaction to that specific interruption but also their subjective experience of it. However, users' receptivity varies with the context as it accounts for their negotiation to handle interruptions in different contexts. 

Consequently, we believe that both attentiveness and receptivity as key dimensions for the design of intelligent notification mechanisms.

\section{Interruptibility Management in Desktop Environments}
\label{sec:im_desktop}

Interruptibility management has attracted the interest of HCI researchers well before the advent of mobile devices. However, interruptions received on the desktop have very specific characteristics. 
In fact, because of their very nature desktops are situated in a constant environment and a user's physical context (such as surrounding people, location and physical activity) does not always change while they are interacting with desktops, whereas mobile devices are carried by users almost everywhere, which makes the physical context of these devices very dynamic. Therefore, interruptibility management for desktop environments is in a sense less complex. In Table~\ref{table:im_desktop} we discuss the summary of the literature in this area along with their key findings.

\subsection{Interruptibility Management by Using Wizard of Oz Approach}

In 1993, Horvitz et al. started the Lumiere project~\cite{Horvitz1999Attention} with the goal of building a probabilistic model for making predictions about users' attention under uncertainty. They employed the Wizard of Oz technique to conduct a series of experiments.\footnote{The so-called Wizard of Oz experimental technique involves a user being observed while operating a system whose functionalities are simulated by a wizard (i.e., a hidden observer)~\cite{WizardOfOz}. Here, the system is made to appear as a real one as the wizard, who is located behind the system or remotely connected to the system, fakes the effect of all functions that are not implemented in the system. This makes people believe that they are interacting with a fully functional system and, thus, enables the investigators to monitor the responses of users to their system without actually implementing it.} 
The data obtained from these experiments was later fed to the manually constructed Bayesian networks to predict the users' need by using different variables such as their background, actions and queries. The constructed Bayesian models were used to obtain a probabilistic distribution for the degree of users' attention~\cite{Horvitz2003Models}.

Opportune moments to interrupt people can be better predicted by considering a wide range of context information. For the first time, in ~\cite{Hudson2003Predicting} Hudson et al. explored the possibility of detecting estimators of human interruptibility by using sensors in order to enhance computer mediated communications in work settings. They chose a Wizard of Oz approach that enabled them to simulate sensing functionalities without implementing actual sensors. 
Using the data collected from these sensors they constructed a series of predictive models that achieved a level of the accuracy in the 75-80\% range. Their study provided evidence that sensor information can be effectively used to estimate human interruptibility. Later, Fogarty et al. improved these models by exploiting additional sensor readings~\cite{Fogarty2005}. 
In~\cite{Fogarty2004Examining, Fogarty2005Examining}, the authors further examined the robustness of the proposed sensor-based approach by conducting experiments in a real world scenario.

\begin{table}[t]
\small
\centering 
\begin{tabular}{ | p{0.3\textwidth} p{0.65\textwidth} | } 
\hline 
\rowcolor{Gray}
\textbf{Study Type} & \textbf{Key Findings} \\ [0.05ex] 
\hline 

\multirow{4}{4cm}{Using Wizard of Oz Approach} & Predicting the level of users' attention by using features of their interaction with the system~\cite{Horvitz1999Attention}.\\
	& Predicting interruptibility by using contextual information obtained through simulated sensors~\cite{Hudson2003Predicting, Fogarty2005, Fogarty2004Examining, Fogarty2005Examining}.\\ 
\hline

\multirow{9}{4cm}{Exploiting Task Phases} & Inference of breakpoints within tasks and exploit this information to deliver interruptions~\cite{Czerwinski2000a, Cutrell2001Notification, Beatty1982, Hoeks1993Pupillary, Nakayama2002Act}. \\
	& Inferring cognitive load through pupillary response and according determining the interruptibility~\cite{Iqbal2004Task, Iqbal2005Towards}. \\
	& Determines the time until which the interruption should be held from being delivered in order to minimize the disruption~\cite{Horvitz2005Deferral}. \\
	& Using system usage features to determine interruptibility~\cite{Fogarty2004Examining, Fogarty2005Examining}. \\
	& Predicting the cost of interruption by exploiting the characteristics of task structure~\cite{Iqbal2006Leveraging}. \\
\hline

\multirow{4}{4cm}{On-the-fly Inference of Interruptibility} & Using ambient sensors to detect user's unavailability for telecommunication~\cite{Begole2004Lilsys}.\\
	& Predicting interruptibility by using contextual information including stream of desktop events, meeting status and conversations~\cite{Horvitz2004}.\\ 
\hline

\end{tabular}
\caption{Studies in the area of interruptibility management in desktop environment.} 
\label{table:im_desktop} 
\end{table}

\subsection{Interruptibility Management by Exploiting Task Phases}

In 1986, Miyata and Norman argued that people are less prone to perceive interruptions as disruptive in some phases of a task compared to other phases~\cite{Miyata1986Psychological}. This suggestion was later investigated by Czerwinski et al.~\cite{Czerwinski2000a} in 2000 and by Cutrell et al.~\cite{Cutrell2001Notification} in 2001. Both studies confirm the insights of Miyata and Norman that perceived disruption varies when an interruption arrives at different phases of the task. Moreover, Czerwinski et al.~\cite{Czerwinski2000a} found that interruptions are perceived as less disruptive when they arrive after the primary task is completed. 

On the other hand, in~\cite{Cutrell2001Notification} Cutrell et al. demonstrated that interruptions arriving at the beginning of a primary task are perceived as less disruptive than interruptions occurring at other phases of a task execution. They suggested that interruptions should be deferred until a user is switching tasks, rather than delivering them immediately. They designed an interface for instant messaging that constantly monitors user actions and infers the different phases of the task completion level as well as the moment when the users switch from one task to another. They also suggested that this information can be used to deliver instant messages at opportune moments. 

Horvitz et al. devised the term ``bounded deferral''~\cite{Horvitz2005Deferral}, which utilizes the concept of deferring the interruption if the user is busy and determines the time until which the interruption should be held from being delivered in order to minimize the disruption cost without losing the value of information due to the delay. The interruption is deferred until a maximum amount of time that is pre-specified by users (known as maximal deferral time) and after this maximal deferral time has elapsed, the alert is triggered immediately even if they still remain busy. They examined the \textit{busy} versus \textit{free} states for 113 users over a period of two consecutive days (i.e., a day for each situation). Participants were provided with a ``Busy Context'' tool that allows them to define when they were busy or free. The analysis of the data showed that users switch from busy to free situations in approximately 2 minutes. Moreover, they demonstrated that medium and low urgency emails can be deferred for three and four minutes respectively. They suggested that bounded-deferral policies can reduce the level of interruption while allowing users to be aware of important information.

As discussed earlier in Section~\ref{sec:cost}, Adamczyk et al. demonstrated through a controlled experiment that the breakpoints within a task are the opportune moments to deliver interruptions~\cite{Adamczyk2004If}. Later, in~\cite{Adamczyk2005Method} they proposed a system that can automatically infer the breakpoints in tasks and exploit this information to deliver interruptions. In order to build this system they leveraged the findings of~\cite{Beatty1982, Hoeks1993Pupillary, Nakayama2002Act}, which suggested that users' mental workload has a statistical correlation with the size of the pupil. They validated this in a human-computer interaction environment by showing pupillary response aligns with the changes in mental workload~\cite{Iqbal2004Task, Iqbal2005Towards}. They built a system that uses a head-mounted eye-tracker for measuring users' pupil size. Finally, they showed that their system was able to infer mental workload for route planning and document editing tasks with an average error of 2.81\% and 2.3\% respectively.

In 2007, Iqbal and Bailey investigated the feasibility of inferring different classes of breakpoints (coarse, medium and fine) while task is performed without using any supplementary hardware resources~\cite{Iqbal2007Understanding}. Data was collected from several participants in the form of event logs and screen interaction videos while they were performing the task. They recruited observers to watch the participants performing the task in order to identify and label breakpoints, their type and explanations for their choice. Using the suggestions reported by Fogarty et al. in~\cite{Fogarty2004Examining, Fogarty2005Examining}, they identified a series of features (such as ``switched to another document'', ``closed an application'', ``completed scroll'', and so on) by analyzing observers' explanations for breakpoints and event logs. Based on these features they constructed statistical models, which were able to predict each type of breakpoints with an average accuracy of 69\% to 87\% (i.e., the percentage of correctly identified breakpoints over the total predicted breakpoints). For each type of breakpoint, different features were computed and also different models were constructed.

Previously, in~\cite{Iqbal2006Leveraging} Iqbal and Bailey investigated whether the cost of interruption can be predicted by exploiting the characteristics of task structure. They posited that the cost of interruption is measured only by the resumption lag (i.e., the time taken by users to resume back to their primary tasks after handling an interruption). More specifically, they tried to test whether interruptions arriving at boundaries of subtasks have low cost of disruption, by using the task structure rather than using head-mounted cameras to measure pupil size as they did in~\cite{Iqbal2007Understanding}. Here, \textit{task structure} indicates the decomposition of a task into a sequence of subtasks. The characteristics of the task structure refers to the level at which the task is broken down into subtasks, the type and the mental load associated with these subtasks~\cite{Card1983Psychology}. They evaluated their approach by conducting experiments on a set of primary tasks. In order to collect the data for estimating the resumption lag on receiving interruptions, participants were interrupted at various boundaries of the task execution. Finally, they constructed a statistical model that could predict the cost of interruption with a 56--77\% accuracy for all tasks.

\subsection{On-the-fly Inference of Interruptibility}
All of the studies discussed above were on an \textit{offline} recognition and construction of models to identify opportune moments for triggering interruptions. Lilsys~\cite{Begole2004Lilsys} and BusyBody~\cite{Horvitz2004} were the first attempts at designing solutions for on-the-fly inference of interruptibility. Both systems were built with custom hardware merely for research projects and focused on interruptions in an office environment.

Lilsys~\cite{Begole2004Lilsys} uses ambient sensors to detect user's unavailability for telecommunication. The system consisted of sensors including sound and motion sensors, phone and door usage inference through the attached wires and keyboard/mouse activity inference. Moreover, it allowed users to manually register their unavailability (if they wanted to) and turn on/off the sensing whenever they wanted. Data is collected passively and an inference about the user's presence and availability is made on detecting a change in any sensor event. Lilsys uses the data from phone, keyboard, mouse, motion and sound detectors for predicting the user's presence. At the same time, the unavailability is predicted by exploiting the data from sound, phone, and door sensors. Both predictors are based on a simple Decision Tree model. The system was installed in an office for around seven months. Participants reported the qualitative improvement in the interruptions but not much reduction in the quantity of interruptions.

BusyBody~\cite{Horvitz2004} is based on an initial training phase during which the system asks users about their interruptibility at random times and it also continuously logs the stream of desktop events, meeting status and conversations. Then, on completion of the training phase it uses the collected information to build predictive models based on Bayesian networks for inferring the cost of interrupting users (i.e., the level of perceived disruption) in real-time. Through a small-scale study, the authors demonstrated that BusyBody is able to make predictions about the cost of interrupting users with an accuracy of 70\%-87\%.

Both Lilsys and BusyBody represent valuable applications of machine learning algorithms for exploiting the contextual information to predict interruptibility. Similarly, Iqbal and Bailey proposed OASIS~\cite{Iqbal2010OASIS} -- a system that detects the breakpoints in users' activity independent of any task by exploiting the streams of application events and user interaction with a computer. This information is used to determine the notification scheduling policies on-the-fly and to deliver notifications accordingly.

\section{Interruptibility Management in Mobile Environments}
\label{sec:im_mobile}

\begin{table}[t]
\small
\centering 
\begin{tabular}{ | p{0.3\textwidth} p{0.65\textwidth} | } 
\hline 
\rowcolor{Gray}
\textbf{Study Type} & \textbf{Key Findings} \\ [0.05ex] 
\hline

\multirow{12}{4cm}{Understanding Interruptibility} & People receive around 60-100 notifications each day~\cite{Sahami2014LargeScale, Mehrotra2015NotifyMe}. \\
	& Reported as disruptive but useful for information awareness~\cite{Iqbal2010Notifications}. \\
	& People handle notifications within a few minutes from their arrival~\cite{Sahami2014LargeScale, Mehrotra2015NotifyMe}. \\
	& People manage interruptibility through the ringer mode configurations~\cite{Chang2015investigating}. \\
	& People stop notifications from a specific applications~\cite{Westermann2015Assessing, Lopez2015Managing}. \\
	& Varying importance is given by people to notifications triggered by different applications~\cite{Sahami2014LargeScale}. \\
	& People have social pressure to quickly respond to notifications~\cite{Pielot2014InSitu}. \\
	& People's receptivity to notifications is influences by the'er context, sender-recipient relationship and notification content~\cite{Mehrotra2016MyPhoneAndMe, Mehrotra2017Understanding}. \\
	& People's receptivity to notifications is influences by their mood~\cite{Mehrotra2017MyTraces}. \\
	& People tend to delete apps that keep sending notifications~\cite{Felt2012Problems}. \\
\hline

\multirow{4}{4cm}{Interruptibility Management by Using Current Activity} & Predicting the level of interruption through calls by using users' activity and calendar properties~\cite{Horvitz2005Bayesphone}.\\
	& Predicting and automatic setting of the phone ringer by exploiting the activity~\cite{Rosenthal2011Using}.\\ 
\hline

\multirow{4}{4cm}{Interruptibility Management by Using the Transition between Activities} & Using transitions between physical activities for delivering mobile notifications~\cite{Ho2005}.\\
	& Using breakpoints during users' interaction with mobile phones for delivering notifications~\cite{Fischer2011Investigating, Okoshi2016Towards}.\\ 
\hline

\multirow{9}{4cm}{Interruptibility Management by Using Contextual Data} & Predicting receptivity by using context information~\cite{InterruptMe2014}. \\
	& Predicting attentiveness to notification by using mobile phone usage features~\cite{Pielot2014Didnt, Dingler2015Attentiveness}. \\
	& Predicting call availability by using the phone usage activity and contextual information~\cite{Pielot2014Callavailability}. \\
	& Predicting receptivity by using the notification content (i.e., information type and relationship with the sender) and context information~\cite{Mehrotra2015NotifyMe}. \\
\hline

\multirow{6}{4cm}{Filtering Irrelevant Information} & Notifications that are uninteresting or irrelevant to people's interests are mostly dismissed~\cite{Sahami2014LargeScale}. \\
	& Predicting receptivity to notifications by the predefined relevant and irrelevant informations classes~\cite{Fischer2010Effects}. \\
	& Predicting users' preference about the types of information they want to receive in specific contexts~\cite{Mehrotra2016PrefMiner}.\\

\hline

\end{tabular}
\caption{Studies in the area of understanding and modeling interruptibility in mobile environment.} 
\label{table:im_mobile} 
\end{table}

When mobile devices first appeared, they were used merely for calling and messaging purposes.  
Later, with the advent of sensing capabilities, these devices have graduated from calling instruments to intelligent and highly personal devices performing numerous functions salient to users' daily requirements. This has provided opportunities to mobile applications to connect users to different information channels and deliver them updates in real-time about a variety of events, ranging from personal messages to traffic alerts and advertisements. 
A variety of studies are done in the area of interruptibility management for mobile phones. In this section we focus on the literature of this area, which is summarized in Table~\ref{table:im_mobile}.

\subsection{Understanding Users' Perception Towards Mobile Notifications}

Studies have found that everyday a much higher number of notifications are triggered on people's mobile phone compared to desktop notifications~\cite{Pielot2014InSitu}. As people always carry their mobile phones along with them, mobile notifications could arrive at anytime and anywhere, which makes mobile interruptions inevitable and more obnoxious compared to desktop interruptions. Thus, managing interruptions in mobile settings has also become a more complex and important task. 

Even though people report that notifications are disruptive, they still like to continue receiving them in order to keep themselves aware of newly available information automatically instead of manually checking~\cite{Iqbal2010Notifications}. People tend to use some simple strategies of their own in order to manage interruptions. In a study with several participants~\cite{Chang2015investigating}, Chang and Tang found that people mostly manage interruptibility through the ringer mode of mobile phones. Findings of another study~\cite{Westermann2015Assessing} suggest that only a few people tend to change notification settings for individual applications, such as stopping notifications from a specific application. Similar findings were reported by Lopez et al. in~\cite{Lopez2015Managing}. The authors of the study also suggested that people want a fine-grained control for how, when and which notifications are delivered to them, which is not present in any mobile platform.

In~\cite{Sahami2014LargeScale} the authors conducted a large-scale data collection involving around 200 million notifications from 40,000 mobile phone subscribers to investigate the user behavior on receiving mobile notifications. They found that even though a plethora of notifications are triggered on mobile phones, people handle most notifications within a few minutes from their arrival. Moreover, by analyzing subjective responses of participants, they found that notification triggered by different types (i.e., categories) of apps are assigned different importance by the users. Their findings suggest that notifications comprising information about individuals and events (such as notifications triggered by messaging applications) are given the most importance.

In another recent study~\cite{Pielot2014InSitu}, Pielot et al. found that most mobile notifications received by people are about personal communication, i.e., they are triggered by messenger and email applications. Through the analysis of subjective responses of participants, the authors demonstrate that the social pressure and shared indicators of availability (such as ``the last time the user was online'') provided by communication applications make people respond to personal messenger notifications more quickly. Moreover, their findings suggest that people tend to feel connected on receiving personal message notifications from their social ties. However, increase in the amount of such notifications could lead to increased stress levels of the recipients. Additionally, the authors found that the information triggered by proactive services are considered as least interesting by the users.

In~\cite{Mehrotra2016MyPhoneAndMe} Mehrotra et al. investigated the role of fine-grained features of a delivered notification and an ongoing task on users' attentiveness and receptivity. They demonstrated that ringer mode, presentation of a notification and engagement with a task all influence users' attentiveness. On the other hand, users' receptivity to notifications is influences by the type of information contained in the notification and the relationship between the sender and the recipient. Their findings also suggest that disruptive notifications are accepted by users only when they contain useful information. Finally, they demonstrated that the perceived disruption and response time to a notification both rely on users' personality. In another study~\cite{Mehrotra2017MyTraces}, Mehrotra et al. demonstrated that people's attentiveness to notifications increases in stressful situations, which results in quicker responses to notifications arriving in that period. At the same time, Pejovic et al.~\cite{Pejovic2015} demonstrate that interruptibility is also significantly influenced by the level of users' engagement with their ongoing task.

In~\cite{Fischer2011Investigating} Fischer et al. investigated the behavior of users on receiving notifications from specific categories of applications. They found that the reaction to notifications is a function of the importance of the corresponding application. In another study~\cite{Fischer2010Effects}, Fischer et al. demonstrated that the importance of a notification depends on how interesting, actionable and relevant the delivered information for the recipient is. Findings of another study~\cite{Felt2012Problems} suggest that applications should not trigger uninteresting and irrelevant notifications as people get annoyed and consider deleting applications that consistently triggers such notifications.

\subsection{Interruptibility Management by Using Current Activity}

In~\cite{Horvitz2005Bayesphone} Horvitz et al. proposed the first interruptibility management tool for predicting the cost of interruptions triggered by phone calls. The system consisted of two Bayesian network models that were trained for predicting users' interruptibility and attendance to meetings, which were listed on their personal calendar. The prediction results of these models were used to calculate the cost of interruption through phone calls. The models were trained by using the sensed activity and calendar properties. 
It is worth remarking that this system was designed solely for managing phone calls, which were the only interruptions generated by mobile devices at that time. 

In ~\cite{Rosenthal2011Using} Rosenthal et al. discussed a personalized approach for predicting the cost of a mobile interruption. They conducted a survey to understand mobile phone users' preferences and interruption cost in different situations. Their results suggest that the cost of interruptions vary across users. For instance, a user might not have any problem in receiving interruptions at work, while another user might consider these as a significant disruption. Therefore, according to their findings, interruptibility management models should be personalized. 
%

\subsection{Interruptibility Management by Using the Transition between Activities}

In~\cite{Ho2005} Ho and Intille explored the use of transitions between physical activities for delivering mobile notifications. More specifically, they conducted an experiment to compare users' receptivity to mobile notifications triggered at times corresponding to activity transition and at other times. Their study was based on the hypothesis that this transition indicates ``self interruption'' as users switch to another activity after its completion and the resistance to interruptions might be lower during such moments. The authors customized a few PDAs by adding two wireless accelerometers in each in order to capture users' physical movement. After using temporal smoothing on activity data, they captured four types of transitions: sitting to standing, standing to sitting, sitting to walking and walking to sitting. Their results show that interruptions delivered at a time corresponding to an activity switch are judged more positively compared to interruptions delivered at random times.

In another study~\cite{Fischer2011Investigating} Fischer et al. investigated the use of naturally occurring breakpoints during users' interaction with mobile phones as opportune moments to deliver mobile notifications. Users were asked to report what were they doing on their phones at the time of notification arrival by means of an ESM questionnaire. These notifications were delivered at random times or after the user has finished a call or finished sending/reading an SMS. Their results suggest that notifications, which were delivered just after the user finished a call or sent/read a text message, have received quicker response.  

In ~\cite{Okoshi2016Towards} Okoshi et al. studied the use of breakpoints within users' interaction with a mobile phone for delivering notifications in order to reduce interruptions and improve users' experience. The authors developed Attelia -- a system for detecting breakpoints in users' interaction with mobile phones and to defer notifications until such a breakpoint occurs. Attelia detects breakpoints during the user's interaction with a mobile phone in real-time, by using the sensors embedded in the phone. Attelia monitors users' interaction with applications and exploit this information in order to detect breakpoints. They used the NASA-TLX questionnaire~\cite{Hart1988} to quantify participants' subjective cognitive load. Based on a controlled study with 37 participants, authors demonstrated that the cognitive load of users (who were more sensitivity to interruptions) is reduced by 46\% by triggering notifications at breakpoints compared to triggering notifications at random times. Later, they conducted an ``in-the-wild'' study involving 30 participants in order to validate their mechanism in a real-world scenario. The results of this ``in-the-wild'' study suggest that Attelia could reduce 33\% of the cognitive load by delivering notifications at detected breakpoints. Moreover, the notifications delivered at breakpoints received a quicker response from users.

\subsection{Interruptibility Management by Using Contextual Data}

Advances in the sensing capabilities of mobile phones have allowed for the monitoring of various context modalities, such as location, physical activity and colocation with other Bluetooth devices (i.e., colocation with other users). Numerous studies have demonstrated the potential of exploiting mobile sensing in order to infer not only numerous aspects of users' physical behavioral patterns~\cite{Lane2010, Lane2011CSN, Mehrotra2014SenSocial} but also their health and emotional states~\cite{Rachuri2010EmotionSense, Lu2012, Lane2011b, Lee2012Affective, Likamwa2013MoodScope, Canzian2015MoodTraces, Mehrotra2016MultiModal}.

Scientists have used various  aspects of the physical context of users, captured via mobile sensors, in order to construct machine learning-based models for predicting interruptibility of users. For example, Pejovic et al. developed InterruptMe~\cite{InterruptMe2014} -- an interruption management library for Android-based mobile devices. InterruptMe uses a mixed method of automated smartphone sensing to collect contextual information and experience sampling to ask users about their interruptibility at different moments. This information is exploited by InterruptMe to construct intelligent interruption models based on a series of machine learning algorithms for interruptibility prediction. 
To evaluate InterruptMe, the authors conducted a two-week study with 20 participants and gathered users' \textit{in-the-wild} contextual information such as location, physical activity, emotional state and the level of engagement with an ongoing task. Their results show that opportune moments for interruptions can be predicted with an average accuracy of 60\% and the reported sentiments towards notification can also be predicted with a precision of 64\% and a recall of 41\%. Moreover, they found that the time taken by users to respond to notifications can be predicted accurately. Finally, they demonstrated that the online learning approach can be used to train models well enough to start making stable predictions within a week. 

In~\cite{Pielot2014Didnt} the authors conducted a survey with 84 users to understand people's reaction towards explicit indicators of their availability shared with others on messaging applications (such as ``the last time the user was online''). Their results show that these indicators create a social pressure on users to respond to the messages but people still see a great value in sharing their attentiveness. However, the authors argued that the shared indicators of availability by messaging applications are weak predictors of recipients' attentiveness. Instead, machine-computed prediction of recipients' attentiveness should be used as a more reliable source. In order to validate their proposed approach, the authors conducted an experiment with 24 participants for a period of two weeks. They developed a mobile application that logs information about  users' context and their actual attentiveness, including application name and the arrival time of notifications, response time (i.e., time taken to handle a notification from its arrival), launching and closing times of messaging applications, phone lock/unlock times and the ringer mode. Using this data they computed 17 features and ranked them based on their entropy. They demonstrated that by using only the top seven features, a machine learning algorithm can construct a model that can predict users' attentiveness with an accuracy of 70.6\%.

In~\cite{Dingler2015Attentiveness} Dingler and Pielot argued that bounded-deferral strategies (i.e., strategies for deferring notifications up to a certain time period in order to reduce disruption caused by it) do not work if users are busy for long time periods. They suggested that a notification might lose its value if it is delayed for too long. Therefore, notifications must only be deferred if the phases of inattentiveness are brief. Based on their hypothesis they conducted a study to investigate whether users' attentiveness to mobile phone notifications can be predicted by using contextual information. Through a passive and continuous sensing approach, they collected phone-usage data from 42 participants for a period of two weeks. They demonstrated that users' attentiveness can be predicted using mobile phone usage with an accuracy of 80\%. Their findings show that users are attentive to mobile notifications for around 12 hours in a day. Also, the periods of users' inattentiveness to mobile notifications are often very short (i.e., 2-5 minutes). 

Another study~\cite{Pielot2014Callavailability} explored the use of phone usage activity and contextual information for predicting users' attentiveness to calls. In order to collect data, the authors developed an application that temporarily mutes the ringer by simply shaking the phone. They logged anonymous data from 418 users corresponding to more than 31000 calls mapped with recipients' context at the time of call arrival. They collected data which is available through the open API calls of the Android platform, such as physical activity, ringer mode, device posture and time since last call. Their results demonstrated that by exploiting these features users' availability for calls can be predicted with 83.2\% accuracy. Moreover, they showed that a personalized model training approach can increase the average accuracy by 87\%. By using only the top five features (including last time the ringer mode was changed, last time the screen was locked/unlocked, current status of screen lock, last time the phone was plugged/unplugged from charging, time since last call) call availability can be predicted with an accuracy of 79.62\%.


In~\cite{Grandhi2009Conceptualizing}, Grandhi et al. argued that interruptibility prediction models often fail to infer the opportune moments to deliver information because they do not consider exploiting the sender-recipient relationship and the type of information. In order to validate this, Mehrotra et al.~\cite{Mehrotra2015NotifyMe} conducted a real-world study for over three weeks and collected around 70,000 instances of notifications from 35 users. More specifically, they exploited users' contextual data, delivered information's type and the sender-recipient relationship for modeling interruptibility. Their results show that the notification receptivity can be predicted with sensitivity and specificity of 70\% and 80\% respectively, which can go 10\% higher for some users. Moreover, the authors demonstrate that their interruptibility prediction model outperforms subjective rules with which users describe their interruptibility.

\subsection{Interruptibility Management by Filtering Irrelevant Information}

Previous studies have found that in order to not miss any newly available important information, people are willing to tolerate mobile notifications to interrupt them~\cite{Iqbal2010Notifications}. However, various mobile applications exploit their willingness by triggering a large number of notifications~\cite{Pielot2014InSitu}. At the same time, some studies have demonstrated that users do not accept all notifications as their receptivity relies on the type and sender of information being delivered~\cite{Mehrotra2015NotifyMe, Mehrotra2016MyPhoneAndMe}. For this reason, notifications that are uninteresting and irrelevant are mostly dismissed (i.e., swiped away without clicking) by users~\cite{Fischer2010Effects, Sahami2014LargeScale}. Some examples include new game invites, app updates, predictive suggestions by recommendation system and marketing emails. At the same time, previous studies have shown continual trigger of such irrrelevant notifications becomes a cause of annoyance for users, which could result in uninstalling the corresponding application~\cite{Felt2012Problems, Sahami2014LargeScale}. These findings suggest that interruptibility management system should take into consideration the relevance and users' interest towards delivered notifications. 


In~\cite{Fischer2010Effects} Fisher et al. examined whether the user's receptivity to a notification is influenced by its content and the time of delivery. In order to understand the role of notification content, they recruited 11 participants and asked them to report their interest for the given 28 categories of content on a 7-point Likert scale. For each participant, the content types rated (by the same participant) as top 3 were considered as ``good content'' and the lowest three were considered as ``bad content''. Moreover, participants were asked to specify the time window during which they were interested in receiving notifications of these types. Participants received six notifications (three each of good and bad content) every day. These notifications were delivered at three opportune times and three inopportune ones. Their results show that users' receptivity is significantly higher for good content compared to bad content. However, there was no significant differences in users' receptivity at opportune and inopportune times. This might suggest that users' receptivity is associated with notification content rather than the time of delivery. Furthermore, they collected the subjective responses of participants to investigate the role of notification content's characteristics for influencing users' receptivity. Their findings suggest that users' interest, entertainment, relevance and the actions required to handle the notification significantly influence their receptivity.


In a recent study~\cite{Mehrotra2016PrefMiner, Mehrotra2017GetMobile}, Mehrotra et al. have presented the design of a solution for managing interruptibility by discovering rules for users' receptivity in different contexts through the analysis of their past interaction with notifications. They first evaluate their approach on the My Phone and Me study's dataset~\cite{Mehrotra2016MyPhoneAndMe} that comprises users' interaction with ``in-the-wild'' notifications. Their analysis demonstrated that, by exploiting only the user's location and the title of a notification, their system could predict with 91\% precision if that notification will be accepted or declined. Furthermore, they funneled their findings into the development of \textit{PrefMiner} -- an Android library for notification management that offers an API to extract rules for the user's receptivity. Interestingly, unlike previous interruptibility studies, Mehrotra et al. not just performed the offline evaluation of their mechanism on the collected data, but they conducted another experiment to carry out an in-the-wild study of the usage of PrefMiner. Their results demonstrated that PrefMiner suggested 179 rules out of which 56.98\% were accepted by users. These rules were able filter out unwanted notifications with 45.81\% accuracy.

\section{Limitations of the State of the Art and Open Challenges}
\label{sec:discussion}

Interruptions are an inevitable part of our daily life. As discussed in this survey, the effects of disruption caused by interruptions occurring at inopportune moments have been studied thoroughly in the past. Numerous studies have been conducted to investigate the effect of interruptions on users' ongoing tasks. More specifically, researchers have found that completion time~\cite{Cutrell2001Notification, Czerwinski2000Instant, Monk2002Attentional}, error rate~\cite{Latorella1998Effects}, and even emotions towards the ongoing tasks~\cite{Adamczyk2004If, Bailey2006Need} are adversely affected by interrupting users at inappropriate moments. However, studies have also provided evidence that in order to not miss any newly available important information, people tolerate some disruption~\cite{Iqbal2010Notifications}.

Since the era of desktops, managing interruptions has been a key theme in Human-Computer Interaction research~\cite{Horvitz1999Attention, Horvitz2004, Adamczyk2005Method, Iqbal2006Leveraging}. With the advent of mobile technologies, the problem of managing interruptibility has become even more pressing as users can now receive notifications anywhere and at anytime. Indeed, numerous research efforts have been carried out with the goal of designing interruptibility management system for mobile environments~\cite{Ho2005, Iqbal2010Notifications, InterruptMe2014, Okoshi2016Towards, Mehrotra2015NotifyMe, Mehrotra2016PrefMiner}. Most of the work on mobile interruptibility emphasizes the exploitation of features that can easily be captured through mobile sensors, such as task phases~\cite{Ho2005, Fischer2011Investigating, Iqbal2010Notifications}, users' context including location and activity~\cite{InterruptMe2014, Pielot2014Didnt, Dingler2015Attentiveness}, and notification content~\cite{Mehrotra2015NotifyMe, Mehrotra2016PrefMiner} to infer the right time to interrupt. More specifically, studies have shown that notifications are considered more positively and received a faster response when delivered while a user switches from one activity to another~\cite{Ho2005, Fischer2011Investigating}. On the other hand, studies have also demonstrated that machine learning algorithms can learn about users' interruptibility by exploiting passively sensed contextual information and notification content~\cite{InterruptMe2014, Mehrotra2015NotifyMe}. Furthermore, contextual information can also be exploited to infer and filter out the irrelevant information from being delivered~\cite{Mehrotra2016PrefMiner}.

Consequently, existing studies have focussed on various challenges concerning the understanding and learning users' behavior in terms of interactions with notifications. However, the characterization of attentiveness and receptivity of users for mobile notifications is still an open problem. We believe that there is still a considerable scope for improvement, for example by exploiting  other physical, social and cognitive factors for modeling users' notification interaction behavior. For instance, more knowledge about users' cognitive context could help the system to reduce the amount of notifications delivered to users when they are stressed~\cite{Mehrotra2017Understanding}. We now summarize some key open questions in the area that must be investigated to build intelligent mechanisms that could effectively trigger the \textit{right} information in a given context. \\

\noindent {\bf Deferring Notifications.} Let us start from a key open question in this area: should we defer a notification if it is not delivered at an opportune moment and for how long? Until now, interruptibility management studies have focused on inferring if the current moment is opportune to deliver notifications or not. We believe that in order to be more effective, these systems should not just predict users' current interruptibility, but if the current time is not an opportune one, it should also \textit{anticipate} the best moment in the nearest future~\cite{Mehrotra2015ESM}. This would enable the overlying application to decide whether the notification should be deferred until the predicted opportune moment or not. Previous studies have proposed mechanisms for anticipating certain types of context modalities such as location~\cite{Nextplace2011}. We believe that similar prediction models can be adopted or developed in order to address this challenge. \\

\noindent {\bf Monitoring Cognitive Context.} Previous approaches for interruptibility management focus on exploiting users' physical context. However, as discussed in this survey, users' interruptibility might also be associated with their cognitive context. There is indeed a need for developing and evaluating mechanisms for automatically capturing the level of users' engagement with the current task, complexity and difficult of execution of the interrupting task, and similar cognitive factors that might influence interruptibility. One of the potential approaches might be to explore the use of affective computing~\cite{Picard1997AffectiveComputing} in order to monitor users' emotional states.  \\

\noindent{\bf Learning ``Good" Behavior: Interruptions for Positive Behavior Intervention.} Until now, all interruptibility studies have focused on learning the observed user behavior associated with the sensed contextual information. However, interruptibility management system can also be considered as a key component for behavior change intervention tools that could help prevent and modify \textit{harmful behavior} of users~\cite{Lathia2013}. In other words, a potential future direction of these systems could also be to gather the knowledge about \textit{good behavior} and exploit it to improve the behavior of users. Such knowledge about \textit{good behavior} for interacting with notifications could potentially be obtained by carrying out a large-scale ESM-based study that can query users about the ideal notification-interaction behavior in their current situation. However, there is an inherent problem related to learning ``good'' versus ``bad'' behavior. The problem is inherent in the fact that a machine learning might not distinguish between a behavior that should be promoted and one that should not. As an example, let us consider a learning component that is able to learn the right moment to interrupt by past experience. If upon receiving a notification a user reads emails on their mobile phone while driving, the notification mechanism should \textit{not} learn this behavior and deliver emails accordingly. Instead, the mechanism should infer that it is a \textit{harmful behavior} to read notifications while driving and try to avoid sending unnecessary emails. Indeed, if the information is critical it should be delivered immediately regardless of the current situation. However, designing such an ideal notification delivery mechanism is extremely difficult and have never been considered in the scope of any interruptibility study. \\

\noindent {\bf Modeling for Multiple Devices.} 
A recent study shows that users? prefer to receive notifications on specific devices based on their situation~\cite{Weber2016Multidevice}. However, none of the previous studies have focused on predicting users' behavior on receiving cross-platform notifications, which are delivered on multiple devices at the same time. In other words, there is a lack of understanding of the features that determine users' receptivity to such notifications on a specific device in a given context. Given the fact that users are surrounded by an increasing number of devices that are able to receive a notifications (such as apps in laptops, mobile phones, wearables, smart television sets and appliances), the design of such mechanism is an open and interesting research area. \\

\noindent {\bf Need for Large-scale Studies.} Almost all studies in the area of interruptibility management are conducted with small samples of the population and for short time periods. At the same time, these studies are often publicized through the network of the researchers performing the studies. This could introduce a bias deriving from the self-selected sample of users and thus the behavior of a certain group of user (within a network) might be different from others. Moreover, the validation of the interruptibility prediction mechanisms in these studies is usually performed in an offline fashion (i.e., the evaluation is performed on the collected data \textit{a posteriori}). For this reason, the results presented in these studies do not have ecological validity as the collected datasets might be biased towards a certain group of population. 
Therefore, we believe that there is a need for large scale studies as well as \textit{in-the-wild} deployments~\cite{rogers2017research, consolvo2017mobile} to guarantee the ecological validity and robustness of the proposed interruptibility prediction mechanisms. We also believe that reproducing these studies in different social context and users' demographics is also essential.

\section{Summary}
\label{sec:summary}

The always-on connectivity of mobile phones and wearables have made them a unique platform to push information in real-time and, thus, they represent a medium for receiving information in an effortless way. However, due to the mobile nature of these devices the inevitable notifications often arrive at inopportune moments. In this article we surveyed the key studies in the area of interruptibility management that demonstrated that such disruptive notifications can adversely affect the ongoing task and affective state of the user. Moreover, with the advent of advanced technologies in mobile phones, this tension is exacerbated as individuals have to deal with a plethora of mobile notifications everyday, some of which are disruptive. Overall, these findings provide evidence for the need of a smart notification mechanism that can reduce the level of disruption by delivering the information at opportune moments.

We have also devoted special attention to the discussion of the state-of-the-art in modeling and anticipating interruptible moments. Desktop interruptibility management mechanisms have mainly focused on task transition phases, where tasks are related to interaction with the device itself. On the other hand, in existing mobile interruptibility studies, anticipatory models predicts users' interruptibility by relying on their physical contextual information sampled through mobile sensors and the content of notifications. Furthermore, we note that apart from handful of interruptibility management mechanisms most models are evaluated only in the offline settings. This is due to the fact that performance of these intelligent mechanisms do not always meet users' expectations in terms of usability and correctness. Indeed, learning the behavior of users for interacting with notifications is not an easy task as it relies on various contextual modalities, both physical and cognitive.

Moreover, we note that all the proposed interruptibility management mechanisms aim at learning the observed user behavior associated with the sensed contextual information and adapt the notification delivery process accordingly. Ideally, such mechanisms should also possess the knowledge about \textit{good behavior} that can be exploited to improve the behavior of users. More in general, intelligent notification systems can be used to deliver intelligent mobile systems.

We believe that further improvements in mobile phones' contextual inference capabilities would also enable researchers to enhance our understanding of users' interaction with notifications. We hope that the critical discussion of interruptibility studies and the overview of the key open challenges presented in this survey would be of valuable help for researchers and practitioners working in this exciting area.


\bigskip
\noindent {\bf ACKNOWLEDGEMENTS}

The authors would like to acknowledge Rami Bahsoon, Russell Beale, Rowanne Fleck, Robert Hendley, Per Ola Kristensson and Veljko Pejovic for the useful comments and discussions about the content of this survey. 

This work was supported by The Alan Turing Institute under the EPSRC grant EP/N510129/1 and at UCL through the EPSRC grant EP/P016278/1.

\balance{}


\end{document}